\documentclass{article}
\usepackage[preprint]{spconf}
\usepackage{amsmath,amsfonts,graphicx}
\usepackage[hyphens]{url}
\usepackage{hyperref}
\usepackage{todonotes}
\usepackage{pgfplots}
\usepackage{tikz}
\usetikzlibrary{shapes}
\usepackage{xcolor}
\usepackage{balance}
\usepackage{multirow}
\usepackage{graphics}

\definecolor{fc}{HTML}{C5E0DC}
\definecolor{cost}{HTML}{F1D4AF}
\tikzset{cost/.style={black,
draw=black,
fill=cost,
rectangle,
rounded corners=10pt,
inner sep=3pt,
minimum height=0.8cm}}
\tikzset{net/.style={black,
draw=black,
fill=fc,
rectangle,
minimum height=1cm}}

\tikzstyle{branch} = [circle,inner sep=0pt,minimum size=1mm,fill=black,draw=black]

\pgfplotsset{my style/.append style={axis x line=middle,axis y line=middle}}

\newtheorem{proposition}{Proposition}

\def\E{{\mathbb E}}
\def\L{{\cal L}}
\def\D{{\cal D}}
\def\S{{\cal S}}

\title{Adversarially Trained Autoencoders for Parallel-Data-Free Voice Conversion}

\name{%
Orhan Ocal$^{\dagger}$,
Oguz H. Elibol$^{\sharp}$,
Gokce Keskin$^{\sharp}$,
Cory Stephenson$^{\sharp}$,
Anil Thomas$^{\sharp}$,
Kannan Ramchandran$^\dagger$%
}
\address{%
\begin{minipage}{8cm}
\centering
$\dagger$University of California at Berkeley\\
Electrical Engineering and Computer Sciences\\
Berkeley, USA%
\end{minipage}
\begin{minipage}{8cm}
\centering
$\sharp$Intel AI Lab\\
Santa Clara, USA%
\end{minipage}
}

\copyrightnotice{\copyright\ IEEE 2019}
\toappear{To appear in {\it Proc.\ ICASSP 2019, May 12-17, 2019, Brighton, United Kingdom}}

\begin{document}
%
\maketitle
\begin{abstract}
We present a method for converting the voices between a set of speakers. 
Our method is based on training multiple autoencoder paths, where there is a single speaker-independent encoder and multiple speaker-dependent decoders. 
The autoencoders are trained with an addition of an adversarial loss which is provided by an auxiliary classifier in order to guide the output of the encoder to be speaker independent.
The training of the model is unsupervised in the sense that it does not require collecting the same utterances from the speakers nor does it require time aligning over phonemes. 
Due to the use of a single encoder, our method can generalize to converting the voice of out-of-training speakers to speakers in the training dataset.
We present subjective tests corroborating the performance of our method. 
\end{abstract}
\section{Introduction}
\label{sec:intro}

Speech signals contain information besides the uttered message; among them are the speech characteristics that pertain to the speaker.
The problem of modifying the speech so that it sounds as if it was uttered by another speaker is known as \emph{voice conversion}~\cite{voice-conversion-sota}.
Voice conversion is usually done by training a model that takes an input from a speaker and transforms it so that it sounds like it was spoken by another speaker.
The training of the model might require \emph{parallel data}, or can be done using \emph{non-parallel data}.
Parallel data refers to recording the same utterance from the speakers, and time-aligning them through dynamic time warping.
On the other hand, non-parallel data refers to recordings from speakers where they speak different utterances, and there is no required time-aligning of signals.

In this paper, we present a method for voice conversion based on autoencoders trained on non-parallel data. 
An autoencoder is an artificial neural network used to learn a representation (encoding) of a given dataset in an unsupervised way~\cite{hinton1994autoencoders}.
This is done by training an encoder network and a decoder network jointly, where the encoder takes an input and gives out a representation (code), and the decoder outputs a reconstruction of the input based on this representation.
The parameters of the encoder and the decoder networks are trained so that the reconstruction matches the input well with respect to a chosen criterion.
The typical scenario where autoencoders are used is to learn an efficient representation of the data by constraining representation to be \emph{smaller} compared to the size of the input~\cite{hinton1994autoencoders}.

Unlike the wide-spread use case of autoencoders, in this work, we use the autoencoder architecture not to find a compact representation of the input, but to learn a representation of the input speech that is \emph{independent} across speakers while still yielding a good reconstruction of the input. 
For this, we use one encoder, multiple decoders (one for each speaker) and one classifier.
The encoder output is guided to a speaker-independent representation in training time by an adversarial loss provided by the classifier.
This classifier network takes as input the output of the encoder (the representation) and tries to identify the speaker.
The encoder-decoder pairs are trained to minimize the reconstruction error while not enabling the classifier to get a good classification accuracy.
In inference time, for performing voice conversion, we feed the speech input to the encoder, and use the decoder of the target speaker.
Because we have a single encoder for all speakers, our algorithm can generalize to converting voices of speakers outside the training set to the speakers' in the training set.

\section{Related Work}
\label{sec:related_work}

Voice conversion algorithms can be divided into two with respect to datasets they require: algorithms which require parallel datasets and the ones that work on non-parallel datasets.

On the side of algorithms based on parallel datasets, the authors of~\cite{stylianou1998continuous} present a method that models the spectral envelope of speech signals by Gaussian mixture models and then fits a conversion function between the source and the target spectral envelops using time-aligned utterances from the two.
Similarly, the authors of~\cite{toda2007voice} propose a conversion method based on the maximum-likelihood estimation of the spectral parameter trajectory instead of working on snapshots.

On the side of parallel-data-free methods, recent work mostly make use of neural networks. 
The authors of~\cite{zhu2017unpaired,zhu2017toward} proposed algorithms to perform domain transfer on images using Generative Adversarial Networks (GANs)~\cite{goodfellow2014generative}.
In order to transfer each sample across domains while preserving the contents of the sample, the authors introduce a new term in the loss function called the \emph{cycle loss}.
This loss tries to enforce that the sample matches the input after it is transferred to the new domain and then back to the old domain.
Voice conversion can be seen as an instance of domain transfer, where the domains correspond to the speakers.
The authors of~\cite{kaneko2017parallel} propose an algorithm to do voice conversion using non-parallel data through the aforementioned GAN architecture trained with the cycle loss.

The closest work to our approach is~\cite{mor2018universal} where the authors propose using adversarially trained autoencoders for translating music across  domains (instruments, genres, and styles).
The authors propose using a universal encoder, and multiple decoders, one for each domain.
The multiple autoencoder paths are trained with an adversarial classification loss.
Our work differs from this related work in the problem domain, the choice of the features tailored for voice conversion on which the loss function is defined, and the network design which is simpler for computationally faster inference.

\section{Adversarially-trained autoencoders for voice conversion}
\label{sec:our_work}

Let $n$ denote the number of speakers whose voices we want to convert to each other, and let $\L$ denote the set of ids of the speakers.
We assume a training dataset consisting of $m$ samples $\S = \{(x_i,l_i)\}_{i=1}^m$, where $x_i$ is the $i$th utterance, and $l_i \in \L$ is the speaker id for the $i$th utterance.

We construct one encoder $E$ and $n$ decoders $\D = \{D_i\}_{i \in \L}$, one for each speaker, and a classifier $C$ that is going to be used for defining the adversarial loss.
Given a reconstruction loss function $f_r$ and a classification loss function $f_c$, the training is done to optimize 
\vspace{-2mm}
\begin{align}
	\min_{E,\D} \max_{C} \sum_{i=1}^m \left[ f_r(x_i,D_{l_i}(E(x_i))) - f_c(l_i,C(E(x_i))) \right].
	\label{eq:cost-fun}
\end{align}
Example choices for the loss functions are an $\ell_p$-norm for the reconstruction loss, that is, $f_r(x,y) = \Vert x-y \Vert_p$, and the cross-entropy loss for the classification, that is, $f_c(l,y) = \log \left({e^{y_l}}/{\sum_{i \in \L} e^{y_i} } \right)$.

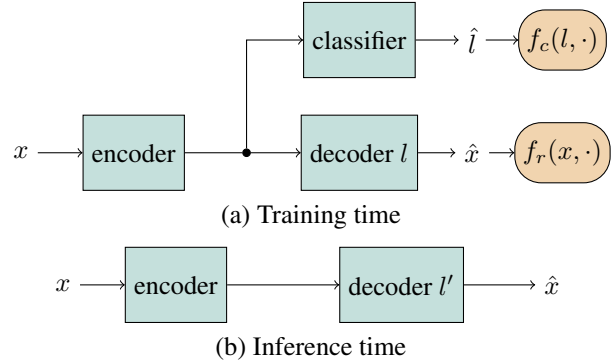
\begin{figure}[bt]
\begin{minipage}[b]{1.0\linewidth}
  \centering
  \centerline{%
  \begin{tikzpicture}
	\node(x) at (0,0) {$x$};
	\node[net] (encoder) at (1.5,0) {encoder};
	\node[branch] (latent) at (3,0) {};
	\node[net] (decoder) at (4.5,0) {decoder $l$};
	\node[net] (classifier) at (4.5,1.5) {classifier};
	\node (hatx) at (6,0) {$\hat{x}$};
	\node (hatl) at (6,1.5) {$\hat{l}$};
	\node[cost] (costclass) at (7.2,1.5) {$f_c(l,\cdot)$};
	\node[cost] (costrecons) at (7.2,0) {$f_r(x,\cdot)$};
	\draw[->] (x) -- (encoder);
	\draw[-] (encoder) -- (latent);
	\draw[->] (latent) -- (decoder);
	\draw[->] (decoder) -- (hatx);
	\draw[->] (classifier) -- (hatl);
	\draw[->] (latent) -- (3,1.5) -- (classifier);
	\draw[->] (hatl) -- (costclass);
	\draw[->] (hatx) -- (costrecons);
	\end{tikzpicture}
  }
  \centerline{(a) Training time}\medskip
\end{minipage}
\begin{minipage}[b]{1.0\linewidth}
  \centering
  \centerline{%
  \begin{tikzpicture}
	\node (x) at (0.5,0) {$x$};
	\node[net] (encoder) at (2,0) {encoder};
	\node[net] (decoder) at (5,0) {decoder $l^\prime$};
	\node (hatx) at (7,0) {$\hat{x}$};
	\draw[->] (x) -- (encoder);
	\draw[->] (encoder) -- (decoder);
	\draw[->] (decoder) -- (hatx);
	\end{tikzpicture}
  }
  \centerline{(b) Inference time}\medskip
\end{minipage}
\caption{Our architecture consists of multiple autoencoder paths, where there is a single encoder and multiple decoders, one for each speaker.  The classifier takes as input the output of the encoder, and it provides the encoder an adversarial loss to guide the representation to be independent of the speaker.}
\label{fig:architecture}
\end{figure}

The intuition behind this particular optimization problem for training is that we want the encoder to learn an embedding that does not carry information with respect to the speaker while still enabling the relevant decoder to reconstruct the speech.
This is facilitated by training the encoder-decoder to result in a large error in the classifier.

There is another way to interpret this cost function in terms of finding representations that have small mutual information with the speaker label.
If the bottleneck of the autoencoder has no mutual information with the label of the speaker, it means that all the information relevant to the speaker has been stripped off the input.
So it would make sense to explicitly minimize this mutual information term as
\begin{align}
	\min_{E,\D} \E \left[ f_r(X,D_{L}(E(X))) \right] + I(L;E(X)).
\end{align}
However, it is hard to calculate and minimize the mutual information.
Recently, a lower-bound based on deep neural networks to approximate the mutual information has been proposed by the authors of~\cite{belghazi2018mine}.
The following proposition shows that, similarly, the classification accuracy can be viewed as a bound on the mutual information.

\begin{proposition}
Let $L$ be the input id of the speaker, $X$ be the speech sample, $Z$ be the representation of $X$ given by the encoder, and $\hat{L}$ be the estimation of speaker id based on $Z$.  Let $p_e = P(L \neq \hat L)$, and $p_e^\ast = \min_f P(L \neq f(Z))$; then
\begin{equation}
\begin{aligned}
	&H(L) - h(p_e) - p_e \log_2( \vert L \vert - 1)\\
	&\leq  I(L;Z) \leq H(L) + \log_2(1 - p_e^\ast).
\end{aligned}	
\label{eq:bounds}
\end{equation}
\end{proposition}

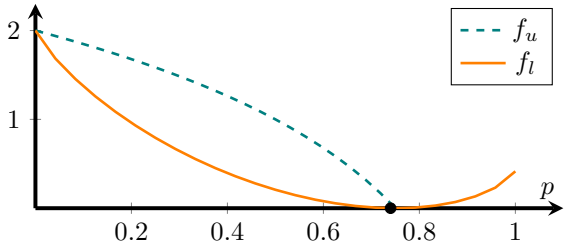
\begin{figure}
	\begin{tikzpicture}
	\begin{axis}[%
	my style,
	axis line style = {line width=1.5pt},
	ymax=2.3,
	xmax=1.1,
	xlabel={$p$},
	height=0.5\linewidth,
	width=\linewidth,
	cycle list name=exotic]
	\addplot+[domain=0:0.75,dashed,no markers,line width=1pt] {log2(4) + log2(1-x)};
	\addplot+[domain=0:1,no markers,line width=1pt] {log2(4) + x*log2(x) + (1-x)*log2(1-x) - x*log2(3)};
	\addplot[mark=*,only marks] coordinates {(0.74,0)};
	\legend{$f_u$,$f_l$}
	\end{axis}
	\end{tikzpicture}
	\caption{An example for the bounds on the mutual information given in equation~\eqref{eq:bounds}. Speaker id $L$ chosen uniformly at random from a set of $4$ speakers. The curve $f_l(p) = H(L) - h(p) - p \log_2( \vert L \vert - 1)$ denotes the lower bound, and the curve $f_u(p) = H(L) + \log_2(1 - p)$ denotes the upper bound on the mutual information $I(L,Y)$.}
	\label{fig:example-bounds}
\end{figure}

The lower bound to the mutual comes from Fano's inequality, and the upper bound comes from manipulating definition of the best classifier; due to space constraints we omit the proof of the proposition.
As can be seen from the proposition, the error probability of any classifier provides a lower bound, and the error probability of the best classifier provides an upper bound on the mutual information.
Hence, in the cost function~\eqref{eq:cost-fun}, the classifier is trained to maximize this lower-bound on the mutual information to get an approximation to it. 
Then, the encoder-decoder pair is trained to minimize this approximation to the mutual information along with the the reconstruction loss.
If the neural network were able to represent the best classifier, and the optimization algorithm were able to find it, then we would be able to get an upper bound on the mutual information as well.
Figure~\ref{fig:example-bounds} shows the lower and upper bounds for the $4$ speakers chosen uniformly at random.

\section{Experiments}
\label{sec:experiments}

\definecolor{conv}{HTML}{C9DBEE}
\tikzset{conv/.style={%
black,
draw=black,
fill=conv,
rectangle,
inner sep=1pt,
text width=1.3cm,
minimum height=1cm}}

\begin{figure}[t]
  \centering
  \begin{tikzpicture}
  	\node[anchor=east, text width=1cm] (desc) at (0,0.8) {Encoder:};
    \node (x) at (0,0) {$x$};
    \node[conv] (fc1) at (1.5,0) {\begin{tabular}{l} $k:3$ \\ $c:256$ \end{tabular}};	
    \node[conv] (fc2) at (3.5,0) {\begin{tabular}{l} $k:3$ \\ $c:256$ \end{tabular}};
    \node[conv] (fc3) at (5.5,0) {\begin{tabular}{l} $k:3$ \\ $c:128$ \end{tabular}};
    \node (z) at (7,0) {$z$};
    \draw[->] (x) -- (fc1);
    \draw[->] (fc1) -- (fc2);
    \draw[->] (fc2) -- (fc3);
    \draw[->] (fc3) -- (z);
  \end{tikzpicture}
  \begin{tikzpicture}
  	\node[anchor=east, text width=1cm] (desc) at (0,0.8) {Decoders:};
    \node (z) at (0,0) {$z$};
    \node[conv] (fc1) at (1.5,0) {\begin{tabular}{l} $k:3$ \\ $c:256$ \end{tabular}};	
    \node[conv] (fc2) at (3.5,0) {\begin{tabular}{l} $k:3$ \\ $c:256$ \end{tabular}};
    \node[conv] (fc3) at (5.5,0) {\begin{tabular}{l} $k:3$ \\ $c:128$ \end{tabular}};
    \node (z) at (7,0) {$\hat{x}$};
    \draw[->] (x) -- (fc1);
    \draw[->] (fc1) -- (fc2);
    \draw[->] (fc2) -- (fc3);
    \draw[->] (fc3) -- (z);
  \end{tikzpicture}
  \begin{tikzpicture}
  	\node[anchor=east, text width=1cm] (desc) at (0,0.8) {Classifier:};
    \node (z) at (0,0) {$z$};
    \node[conv] (fc1) at (1.5,0) {\begin{tabular}{l} $k:3$ \\ $c:256$ \end{tabular}};	
    \node[conv] (fc2) at (3.5,0) {\begin{tabular}{l} $k:3$ \\ $c:256$ \end{tabular}};
    \node[conv] (fc3) at (5.5,0) {\begin{tabular}{l} $k:3$ \\ $c:4$ \end{tabular}};
    \node (z) at (7,0) {$\hat{l}$};
    \draw[->] (x) -- (fc1);
    \draw[->] (fc1) -- (fc2);
    \draw[->] (fc2) -- (fc3);
    \draw[->] (fc3) -- (z);
  \end{tikzpicture}
  \vskip 6px
  \caption{Architectures of the encoder, decoders and classifier.  Each block represents convolution followed by instance normalization an ReLU. The convolution kernel size is denoted by $k$, and the number of output channels is denoted by $c$.}
  \label{fig:layers}
\end{figure}
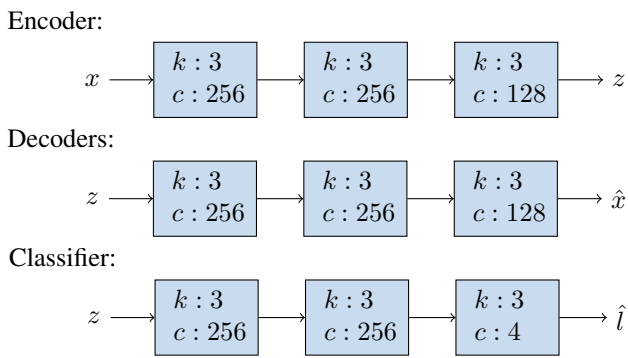

As numerical experiments we select five speakers, two female (denoted with F1 and F2) and three male (denoted with M1, M2 and MX), from English multi-speaker corpus of CSTR voice cloning toolkit~\cite{vctk}.
We use four of these speakers (F1, F2, M1, and M2) in training the neural networks, and set aside one of the speakers (MX) for evaluating the performance for conversion of voice from out-of-training speakers.

We train multiple autoencoder paths on the mel-frequency spectrogram magnitudes of speech.
The parameters for the spectrograms are as follows: the number of FFT bins is $1024$, the hop length between consecutive windows is $256$, number of mel-spaced filters is $128$, and minimum and maximum frequencies of input signal considered are $40$ Hz and $8000$ Hz respectively.
The reconstruction loss, $f_r$, is chosen to be $\ell_1$ loss, and the classification loss, $f_c$, is chosen to be the cross-entropy loss.
The encoder, decoder and classifier are all three layer convolutional neural networks, where the convolutions are $1$-dimensional have kernel size of $3$.
The number of hidden channels for each network is $256$, and the representation size (code length) is chosen to be $128$.
We use instance normalization~\cite{huang2017arbitrary} before each non-linearity, and use ReLU as the activation functions.
Fig.~\ref{fig:layers} shows an illustration of the architecture.
To complete the loop on voice conversion, after the conversion of spectrograms using the neural network, we use Griffin-Lim's algorithm~\cite{griffin1984signal} to construct the audio signal similar to the way Tacotron~\cite{wang2017tacotron} does.

\begin{figure*}[htb]
\centering
\begin{minipage}[b]{0.33\linewidth}
  \centering
  \centerline{%
  \includegraphics[width=\linewidth]{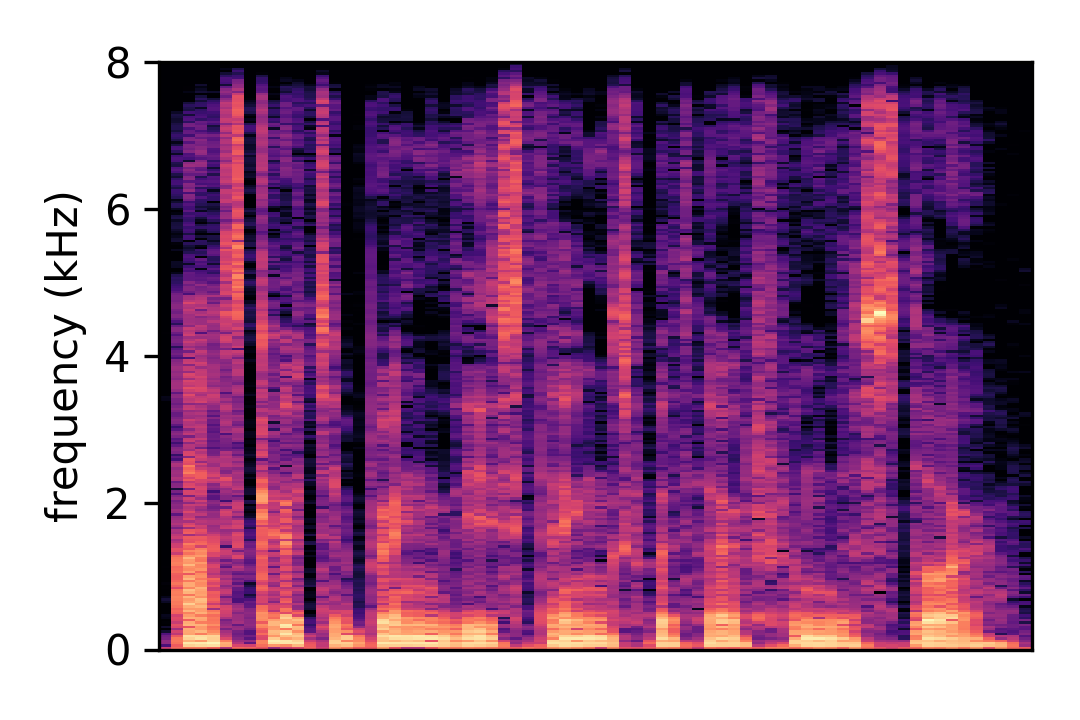}
  }
  \vspace{-3mm}
  \centerline{(a) Input}\medskip
\end{minipage}
\begin{minipage}[b]{0.33\linewidth}
  \centering
  \centerline{%
  \includegraphics[width=\linewidth]{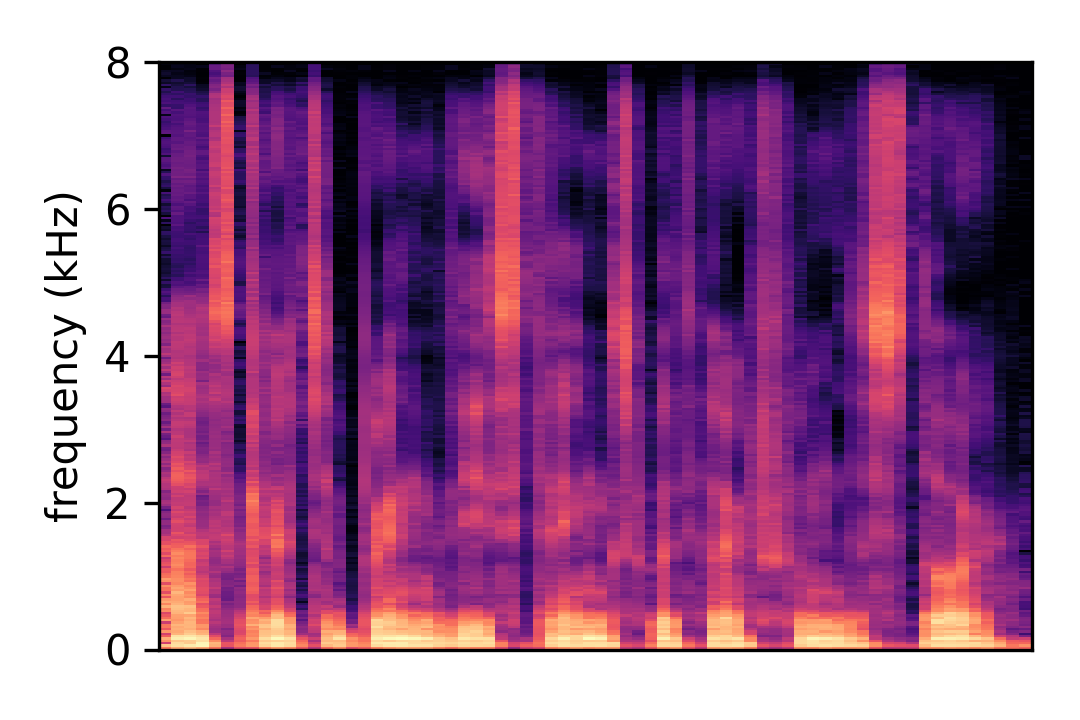}
  }
  \vspace{-3mm}
  \centerline{(b) Reconstruction}\medskip
\end{minipage}
\begin{minipage}[b]{0.33\linewidth}
  \centering
  \centerline{%
  \includegraphics[width=\linewidth]{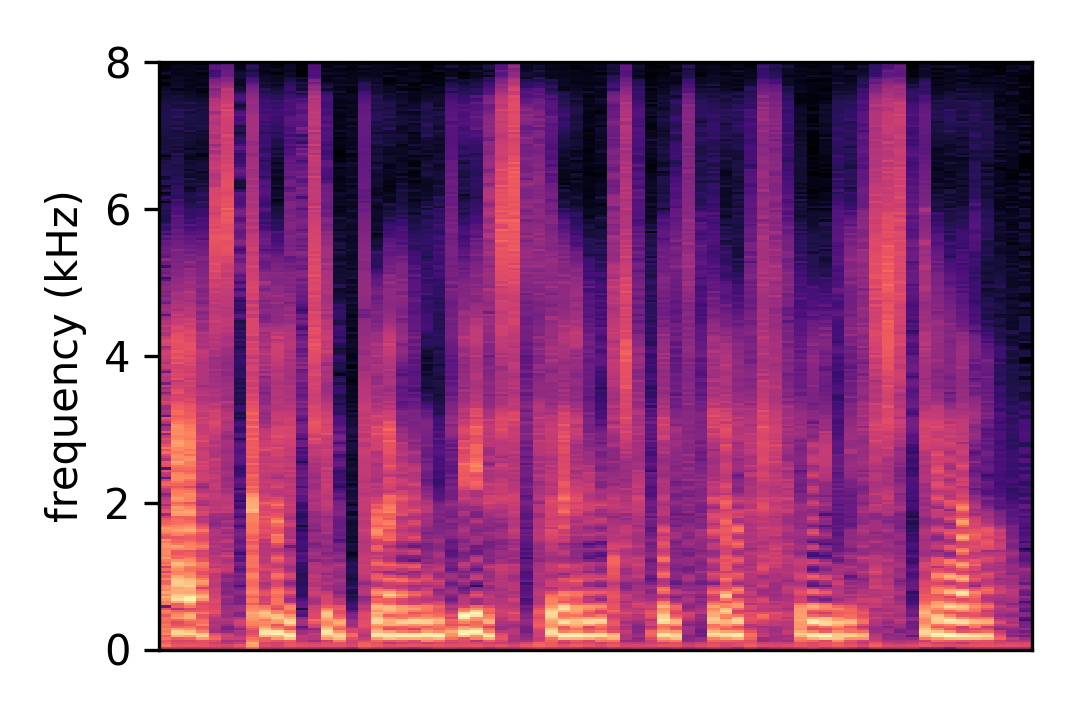}
  }
  \vspace{-3mm}
  \centerline{(c) Conversion}\medskip
\end{minipage}
\vspace{-9mm}
\caption{
	Spectrograms of input speech, reconstruction and voice conversion.
	Input is an utterance by a male speaker and the target is a female speaker who has a higher pitched voice compared to the input speaker.
	We observe that the converted voice has components at higher frequencies compared to the input which aligns with the characteristics of the target speaker.
}
\label{fig:voice-conversion-1}
\end{figure*}

As an example for converted voice spectrogram, we look at converting an input sample from one of the male speakers (M2) to the voice of one of the female speakers (F1). 
Fig.~\ref{fig:voice-conversion-1} shows the input spectrogram, spectrogram of the reconstructed speech and the spectrogram after voice conversion.
As can be seen from the figure, the reconstruction resembles the input spectrogram, and the spectrogram of the converted voice has components at higher frequencies compared to the input's which aligns with the target speaker's voice characteristics.

\renewcommand{\arraystretch}{1.5}
\begin{table}[tb]
\resizebox{\columnwidth}{!}{%
\begin{tabular}{c|c|c|c|c|c|}
\multicolumn{2}{c}{} & \multicolumn{4}{c}{target speaker} \\ 
\cline{3-6}
\multicolumn{2}{c}{} & 
\multicolumn{1}{c}{F1} &  \multicolumn{1}{c}{F2} & 
\multicolumn{1}{c}{M1} & \multicolumn{1}{c}{M2} \\ 
\cline{3-6}
\multirow{5}{*}{\rotatebox{90}{source speaker}} 
& F1  & $-$ & $0.500 \pm 0.224$  & $0.825 \pm 0.069 $ &  $0.800 \pm 0.080$  \\
\cline{3-6}
& F2  & $ 0.567 \pm 0.128$ & $-$  &  $0.756 \pm 0.091$ & $ 0.750 \pm 0.087$  \\ 
\cline{3-6}
& M1  & $0.846 \pm 0.063$ & $0.818 \pm 0.074$   & $-$ & $0.776 \pm 0.064$  \\ 
\cline{3-6}
& M2  &  $ 0.914 \pm 0.067$   & $0.856 \pm 0.074$   & $0.675 \pm 0.105$ & $-$  \\ 
\cline{3-6}
& MX  &  $0.867 \pm 0.088$   & $ 0.900 \pm 0.072$   & $0.607 \pm 0.083$ & $ 0.767 \pm 0.089$  \\ 
\cline{3-6}
\end{tabular}
}
\caption{AB test results.}
\label{tab:ab}
\end{table}

\begin{table}[tb]
\resizebox{\columnwidth}{!}{%
\begin{tabular}{c|c|c|c|c|c|}
\multicolumn{2}{c}{} & \multicolumn{4}{c}{target speaker} \\ 
\cline{3-6}
\multicolumn{2}{c}{} & 
\multicolumn{1}{c}{F1} &  \multicolumn{1}{c}{F2} & 
\multicolumn{1}{c}{M1} & \multicolumn{1}{c}{M2} \\ 
\cline{3-6}
\multirow{5}{*}{\rotatebox{90}{source speaker}} 
& F1  & $-$ & $0.651 \pm 0.103$  & $0.872 \pm 0.069$ &  $0.868 \pm 0.110$  \\
\cline{3-6}
& F2  & $0.650 \pm 0.151$   & $-$  &  $0.912 \pm 0.069$ & $0.860 \pm 0.075$  \\ 
\cline{3-6}
& M1  & $0.855 \pm 0.081$   & $0.826 \pm 0.082$   & $-$ & $0.776 \pm 0.069$  \\ 
\cline{3-6}
& M2  &  $0.750 \pm 0.137$   & $0.840 \pm 0.077$   & $0.634 \pm 0.081$ & $-$  \\ 
\cline{3-6}
& MX  &  $0.776 \pm 0.110$   & $0.833 \pm 0.063$   & $0.653 \pm 0.064$ & $0.800 \pm 0.179$  \\ 
\cline{3-6}
\end{tabular}
}
\caption{ABX test results.}
\label{tab:abx}
\end{table}

\begin{table}[tb]
\resizebox{\columnwidth}{!}{%
\begin{tabular}{c|c|c|c|c|c|}
\multicolumn{2}{c}{} & \multicolumn{4}{c}{target speaker} \\ 
\cline{3-6}
\multicolumn{2}{c}{} & 
\multicolumn{1}{c}{F1} &  \multicolumn{1}{c}{F2} & 
\multicolumn{1}{c}{M1} & \multicolumn{1}{c}{M2} \\ 
\cline{3-6}
\multirow{5}{*}{\rotatebox{90}{source speaker}} 
& F1  & $-$ & $2.371 \pm 0.232$ & $2.312 \pm 0.195$ & $2.138 \pm 0.192$ \\
\cline{3-6}
& F2  & $1.850 \pm 0.205$ & $-$ & $1.712 \pm 0.180$	& $1.750 \pm 0.160$ \\
\cline{3-6}
& M1  & $1.588 \pm 0.160$ & $1.754 \pm 0.126$ & $-$ & $2.327 \pm 0.201$ \\
\cline{3-6}
& M2  & $2.140 \pm 0.159$ &	$2.020 \pm 0.164$ & $2.689 \pm 0.218$ & $-$ \\
\cline{3-6}
& MX  & $2.670 \pm 0.182$ & $2.422 \pm 0.190$ & $2.650 \pm 0.199$ & $2.577 \pm 0.19$ \\
\cline{3-6}
\end{tabular}
}
\caption{MOS.}
\label{tab:mos}
\end{table}

In order to get quantitative performance metrics, we conducted subjective tests on Amazon Mechanical Turk~\cite{amazons-mechanical-turk}.
We report the results of three types of tests: (1) AB testing, (2) ABX testing, and (3) mean opinion scores (MOS).

The first of the subjective experiments is AB tests.
In these tests, the listeners are given two utterances to compare.
The first sample (A) is a recording by either the source or the target speaker chosen uniformly at random, and the second one (B) is an utterance converted to the voice of the target speaker from the source.
In order to not bias the listener based on the contents of the utterances, we use different sentences for the given two samples.
The subjects are asked to identify if the two samples are spoken by the same speaker or not.
We call it a success if (1) the first sample (A) was a recording from the target and the subject said the two samples were from the same speaker, or (2) the first sample (A) was a recording from the source and the subject said the two samples were from different speakers.

Table~\ref{tab:ab} shows the frequency of success in the experiments along with $95\%$ confidence intervals.
As can be seen from the table, our algorithm achieves to change the source voice in the correct direction.
In particular, the conversion of voice between speakers of different genders seem to be consistently perceived successfully by the subjects.
An interesting observation is that converting the voice of an out-of-training speaker (given in the last row of the table) performs similarly with in-training speakers, which means that the encoder can generalize to out-of-training speakers.

The second subjective experiment is ABX tests.
Here, the listeners are given three samples.
The first two samples (A and B) are recordings from the source and target speakers (order is randomized), and the third sample (X) is an utterance converted from the source speaker to the voice of the target speaker.
As in AB tests, we use different sentences for the given three samples in order  not to bias the listener based on the contents of the utterances.
The subjects are asked to choose which of the first two samples' voice (A's or B's) is the third sample's voice is closer to.
We call it a success if the subject chooses the sample recorded from the target speaker.

Table~\ref{tab:abx} shows the frequency of success in the experiments along with $95\%$ confidence intervals.
Again, our algorithm changes the source voice in the correct direction, in particular, for converting between speakers with different genders.
Similar to AB tests, the voice of an out-of-training speaker (last row of the table) performs similarly with in-training speakers.

The last of the subjective tests is MOS.
Here, the subjects were given converted speech samples, and were asked to rate the quality of the sample.
The quality scale used was Absolute Category Rating (ACR)~\cite{itu1999subjective}, that is, integers with correspondences: 1-bad, 2-poor, 3-fair, 4-good, 5-excellent.

Table~\ref{tab:mos} shows the MOS along with $95\%$ confidence intervals.
Even though, our algorithm changes the source voice in the correct direction as was shown with the AB and ABX tests, the converted sample has artifacts that a listener can notice as is evident from the MOS.
A large part of it is because small inconsistencies in the spectrogram magnitudes can result in significant artifacts in the output when Griffin-Lim's algorithm is used.
This behavior was also identified by the authors of~\cite{arik2017deep} who propose training a separate deep neural network that takes a spectrogram magnitude and outputs audio signals to reduce artifacts.
Using such more elaborate approaches to reduce artifacts is part of ongoing research.

\section{Discussion}
\label{sec:discussion}

We presented a method for voice conversion using neural networks trained on non-parallel data. 
The method is based on an training multiple autoencoder paths where there is a single speaker-independent encoder and multiple speaker-dependent decoders. 
The autoencoder paths are trained to minimize the reconstruction error and an adversarial cost that tries to make the output of the encoder carry no information with respect to the speaker id.
The training is unsupervised in the sense that we do not require parallel speech dataset from the speakers. 
We evaluated our method on a subset of speakers from the VCTK dataset.
Qualitatively, we observe that the converted spectrograms carry characteristics of the spectrograms of the target speaker. 
The results of subjective tests corroborate our algorithm's voice conversion performance.
Although our algorithm can convert the voice of the source speaker in the direction of the target, we observe that reconstructed audio has some artifacts.
Reducing these artifacts is ongoing work.

\newpage
\balance
\bibliographystyle{IEEEbib}
\bibliography{refs}

\begin{thebibliography}{10}

\bibitem{voice-conversion-sota}
E~Moulines and Y~Sagisaka,
\newblock ``Voice conversion: State-of-the-art and perspectives,''
\newblock {\em Speech Communication}, vol. 16, no. 2, pp. 125--126, Feb 1995.

\bibitem{hinton1994autoencoders}
Geoffrey~E Hinton and Richard~S Zemel,
\newblock ``Autoencoders, minimum description length and helmholtz free
  energy,''
\newblock in {\em Advances in neural information processing systems}, 1994, pp.
  3--10.

\bibitem{stylianou1998continuous}
Yannis Stylianou, Olivier Capp{\'e}, and Eric Moulines,
\newblock ``Continuous probabilistic transform for voice conversion,''
\newblock {\em IEEE Transactions on speech and audio processing}, vol. 6, no.
  2, pp. 131--142, 1998.

\bibitem{toda2007voice}
Tomoki Toda, Alan~W Black, and Keiichi Tokuda,
\newblock ``Voice conversion based on maximum-likelihood estimation of spectral
  parameter trajectory,''
\newblock {\em IEEE Transactions on Audio, Speech, and Language Processing},
  vol. 15, no. 8, pp. 2222--2235, 2007.

\bibitem{zhu2017unpaired}
Jun-Yan Zhu, Taesung Park, Phillip Isola, and Alexei~A Efros,
\newblock ``Unpaired image-to-image translation using cycle-consistent
  adversarial networks,''
\newblock {\em arXiv preprint}, 2017.

\bibitem{zhu2017toward}
Jun-Yan Zhu, Richard Zhang, Deepak Pathak, Trevor Darrell, Alexei~A Efros,
  Oliver Wang, and Eli Shechtman,
\newblock ``Toward multimodal image-to-image translation,''
\newblock in {\em Advances in Neural Information Processing Systems}, 2017, pp.
  465--476.

\bibitem{goodfellow2014generative}
Ian Goodfellow, Jean Pouget-Abadie, Mehdi Mirza, Bing Xu, David Warde-Farley,
  Sherjil Ozair, Aaron Courville, and Yoshua Bengio,
\newblock ``Generative adversarial nets,''
\newblock in {\em Advances in neural information processing systems}, 2014, pp.
  2672--2680.

\bibitem{kaneko2017parallel}
Takuhiro Kaneko and Hirokazu Kameoka,
\newblock ``Parallel-data-free voice conversion using cycle-consistent
  adversarial networks,''
\newblock {\em arXiv preprint arXiv:1711.11293}, 2017.

\bibitem{mor2018universal}
Noam Mor, Lior Wolf, Adam Polyak, and Yaniv Taigman,
\newblock ``A universal music translation network,''
\newblock {\em arXiv preprint arXiv:1805.07848}, 2018.

\bibitem{belghazi2018mine}
Ishmael Belghazi, Sai Rajeswar, Aristide Baratin, R~Devon Hjelm, and Aaron
  Courville,
\newblock ``Mine: mutual information neural estimation,''
\newblock {\em arXiv preprint arXiv:1801.04062}, 2018.

\bibitem{vctk}
Junichi Yamagishi,
\newblock ``{English multi-speaker corpus for CSTR voice cloning toolkit},''
  \url{http://homepages.inf.ed.ac.uk/jyamagis/page3/page58/page58.html }, 2012.

\bibitem{huang2017arbitrary}
Xun Huang and Serge~J Belongie,
\newblock ``Arbitrary style transfer in real-time with adaptive instance
  normalization.,''
\newblock in {\em ICCV}, 2017, pp. 1510--1519.

\bibitem{griffin1984signal}
Daniel Griffin and Jae Lim,
\newblock ``Signal estimation from modified short-time fourier transform,''
\newblock {\em IEEE Transactions on Acoustics, Speech, and Signal Processing},
  vol. 32, no. 2, pp. 236--243, 1984.

\bibitem{wang2017tacotron}
Yuxuan Wang, RJ~Skerry-Ryan, Daisy Stanton, Yonghui Wu, Ron~J Weiss, Navdeep
  Jaitly, Zongheng Yang, Ying Xiao, Zhifeng Chen, Samy Bengio, et~al.,
\newblock ``Tacotron: Towards end-to-end speech synthesis,''
\newblock {\em arXiv preprint arXiv:1703.10135}, 2017.

\bibitem{amazons-mechanical-turk}
Michael Buhrmester, Tracy Kwang, and Samuel~D Gosling,
\newblock ``Amazon's mechanical turk: A new source of inexpensive, yet
  high-quality, data?,''
\newblock {\em Perspectives on psychological science}, vol. 6, no. 1, pp. 3--5,
  2011.

\bibitem{itu1999subjective}
P~ITU-T~RECOMMENDATION,
\newblock ``Subjective video quality assessment methods for multimedia
  applications,''
\newblock 1999.

\bibitem{arik2017deep}
Sercan Arik, Gregory Diamos, Andrew Gibiansky, John Miller, Kainan Peng, Wei
  Ping, Jonathan Raiman, and Yanqi Zhou,
\newblock ``Deep voice 2: Multi-speaker neural text-to-speech,''
\newblock {\em arXiv preprint arXiv:1705.08947}, 2017.

\end{thebibliography}

\end{document}